\documentclass[superscriptaddress,twocolumn,floatfix]{revtex4}

\usepackage{graphicx,graphics,epsfig}   
\usepackage{amsmath,amsfonts,amssymb}    

\newcommand{\be}{\begin{equation}}
\newcommand{\ee}{\end{equation}}
\newcommand{\ba}{\begin{eqnarray}}
\newcommand{\ea}{\end{eqnarray}}
\newcommand{\ban}{\begin{eqnarray*}}
\newcommand{\ean}{\end{eqnarray*}}

\newcommand{\ket}[1]{\mbox{$ | #1 \rangle $}}
\newcommand{\bra}[1]{\mbox{$ \langle #1 | $}}

\begin{document}


\title{Closing the detection loophole in Bell experiments using qudits}

\author{Tam\'as V\'ertesi}
\affiliation{ Institute of Nuclear Research of the Hungarian Academy of Sciences
H-4001 Debrecen, P.O. Box 51, Hungary}
\author{Stefano Pironio}
\affiliation{Group of Applied Physics, University of Geneva, CH-1211 Geneva 4, Switzerland}
\author{Nicolas Brunner}%
\affiliation{H.H. Wills Physics Laboratory, University of Bristol, Bristol, BS8 1TL, United Kingdom}

\date{\today}

\begin{abstract}
We show that the detection efficiencies required for closing the detection loophole in Bell tests can be significantly lowered using quantum systems of dimension larger than two. We introduce a series of asymmetric Bell tests for which an efficiency arbitrarily close to $1/N$ can be tolerated using $N$-dimensional systems, and a symmetric Bell test for which the efficiency can be lowered down to $61.8\%$ using four-dimensional systems. Experimental perspectives for our schemes look promising considering recent progress in atom-photon entanglement and in photon hyperentanglement.
\end{abstract}

\maketitle
Quantum theory predicts that measurements on separated entangled systems will produce outcome correlations that are not locally causal, as implied by the violation of Bell inequalities \cite{bell64}. Bell inequality violations have been confirmed in numerous experiments, providing strong indication
that nature is non-local. However, imperfections in such experiments open various loopholes that could in principle be exploited by a locally causal model to reproduce the experimental data. Given the far-reaching significance of non-locality, it appears highly desirable to perform a Bell experiment free of any loopholes. While promising proposals have been made recently \cite{Irvine}, accompanied by significant experimental progress \cite{matsukevich}, a loophole-free Bell experiment is still missing.

There are two basic requirements for a loophole-free Bell experiment. First, a space-like separation between the observers is necessary to ensure that no sub-luminal signal can propagate between the particles. Failure to satisfy this condition is known as the locality loophole. Second, the particle detection efficiency must be larger than a certain level (usually high), otherwise undetected events can be exploited by a local model to reproduce the quantum statistics \cite{pearle,gisingisin}. Failure to satisfy this condition is known as the detection loophole.
All experiments performed so far suffer from at least one of the two above loopholes. Experiments carried out on atoms \cite{rowe,matsukevich} could close the detection loophole, but are unsatisfactory from the locality point of view. Photonic experiments, on the other hand, could close the locality loophole \cite{aspect}, but current photon detection efficiencies are still too low for closing the detection loophole.

Closing the detection loophole is also crucial from the perspective of quantum information processing applications based on the non-local correlations of entangled states, such as communication complexity \cite{cc}, device-independent cryptography \cite{DevIndep} or state tomography \cite{st}. Indeed, if the detection loophole is not closed, then the data produced in the experiment could as well have been produced by classical means, and thus does not provide any advantage.

For the Clauser-Horne-Shimony-Holt (CHSH) Bell inequality \cite{chsh}, the threshold detection efficiency is $82.8\%$ for a maximally entangled qubit pair and can be lowered down to $66.7\%$ using partially entangled states, as shown by Eberhard \cite{Eberhard}. It appears however very difficult to find Bell inequalities that can tolerate lower efficiencies than CHSH in the bipartite case. For qubits, only marginal improvements have been reported \cite{zoology,I4422,pal09}. Massar has shown that higher dimensional systems can tolerate a detection efficiency that decrease exponentially with the dimension $d$, but this result is of limited practical interest since an improvement over CHSH is obtained only for $d\gtrsim 1600$ \cite{massar}. Up to now, no practical bipartite Bell test was known to tolerate a detection efficiency lower than Eberhard's limit of $66.7\%$.

More recently, the detection loophole has also been studied in an asymmetric configuration \cite{CabelloDetLoop,AsymDetLoop}, inspired from atom-photon entanglement. Since atomic measurements are very efficient ($\eta \approx 1$), a much lower efficiency, compared to the symmetric case, can be tolerated for the photon -- as low as $43\%$ using qubits and a three-setting Bell inequality \cite{AsymDetLoop}.

Here, we show that (low dimensional) qudits offer a significant advantage over qubits for closing the detection loophole. For asymmetric Bell tests, we show that an efficiency arbitrarily close to $1/N$ can be tolerated using $N$-dimensional systems and a family of $N$-setting Bell inequalities introduced by Collins and Gisin \cite{dan}. Our construction is optimal in the sense that a Bell test with $N$ measurement settings cannot tolerate a detection efficiency smaller than $1/N$. In the symmetric scenario, we show that an efficiency as low as $61.8\%$ can be tolerated using four-dimensional states and a four-setting Bell inequality introduced in \cite{I4422}. To the best of our knowledge, these findings improve significantly over all results in the literature for bipartite Bell tests with a reasonable number of measurement settings and dimensions. Moreover, the prospects of experimental implementations look promising considering recent experimental progress in atom-photon entanglement \cite{inoue09} and in photon hyperentanglement \cite{kwiat97,hyperentanglement,hyperMartini}.

Our constructions also provide simple examples of dimension witnesses \cite{dimH,vertesi08}, i.e. Bell-type inequalities which yield a lower bound on the Hilbert space dimension necessary to produce certain quantum correlations.

\emph{Preliminaries.} We consider a Bell-type scenario in which two distant parties, Alice and Bob, can choose among $N$ measurement settings with binary outcomes. Alice's measurement settings are denoted $A_x$ with $x \in \{1,...,N\}$ and her output bit is denoted $a \in \{+1,-1\}$; similarly for Bob we have $B_y$ with $y \in \{1,...,N\}$ and $b \in \{+1,-1\}$. The experiment is characterized by the set of joint probabilities $P(A_x=a,B_y=b)$ to get outcomes $a$ and $b$ when $A_x$ and $B_y$ are measured. It is easily seen that all these probabilities are determined by the following subset of probabilities: $P(A_xB_y)\equiv P(A_x=1,B_y=1)$, $P(A_x)\equiv P(A_x=1)$, and $P(B_y)\equiv P(B_y=1)$, which it is thus sufficient to consider.

The detection efficiencies for Alice and Bob are denoted $\eta_A$ and $\eta_B$. We will study the symmetric configuration, where $\eta\equiv\eta_A=\eta_B$, as in standard photonic experiments, as well as the asymmetric configuration, where $\eta_A=1$ and $\eta_B<1$, inspired from atom-photon entanglement experiments.

\emph{Asymmetric case.} We now introduce a family of Bell tests that involve entangled states whose local Hilbert space dimension is $N$ and which tolerate a detector efficiency arbitrarily close to $1/N$. Our construction is based on a family of Bell inequalities introduced in \cite{dan}, which are given by $I_{NN22}\leq 0$, where
\ba
\nonumber\label{ineq} I_{NN22} &=& -P(A_1) - \sum_{y=2}^N{P(B_y)} + \sum_{y=1}^N{P(A_1,B_y)} \\
& &+ \sum_{x=2}^N{P(A_x,B_x)} - \sum_{1 \leq y< x \leq N}P(A_x,B_y)
\ea
Note that we have written these inequalities in a different form than the one used in \cite{dan}.

Because of the limited efficiency of his detector, Bob does not always obtain a conclusive result. In order to close the detection loophole, one must ensure that the whole set of data, including inconclusive results, violates a Bell inequality. To take into account inconclusive events, we simply choose here that Bob outputs ``-1" in case of non-detection. Thus Bob's output is still binary and the above inequalities can be used. The measurement outcome probabilities are however modified according to $P(A_x,B_y)\rightarrow \eta_ B P(A_x,B_y)$, $P(A_x)\rightarrow P(A_x)$, and $P(B_y)\rightarrow \eta_ B P(B_y)$. Introducing these expressions in (\ref{ineq}) and dividing by $\eta_B$, we obtain the modified (efficiency-dependent) Bell inequalities $I_{NN22}(\eta_{B})\leq0$, where
\ba
\label{ineq2} I_{NN22}(\eta_{B}) & = & I_{NN22}-\frac{1-\eta_{B}}{\eta_{B}}P(A_1).
\ea
A violation of the modified inequality $I_{NN22}(\eta_{B})\leq 0$ implies that the original Bell inequality $I_{NN22}\leq 0$ can tolerate a detection efficiency of $\eta_{B}$ for Bob's detector.

We now give an explicit entangled state and quantum measurements that violate the inequality $I_{NN22}(\eta_{B})\leq0$ when $\eta_{B}>\frac{1}{N}$. Note that our construction is optimal in the sense that no violation is possible if $\eta_B\leq\frac{1}{N}$. Indeed, any Bell test with $N$ measurement settings admits a simple local model when $\eta_B\leq \frac{1}{N}$ \cite{mp03}.

The quantum state is defined in $\mathbb{C}^N\otimes\mathbb{C}^N$ and given in the Schmidt form as
\begin{equation}
\ket{\psi_\epsilon}=\sqrt{\frac{1-\epsilon^2}{N-1}}\left(\sum_{k=1}^{N-1}{\ket{k}\ket{k}}\right)+\epsilon\ket{N}\ket{N},
\label{state}
\end{equation}
with $\epsilon\in [0, 1]$. The measurement operators for Alice can be written as $A_x=A_x^+-A_x^-$, where $A_x^+$ is the projector on the $+1$ subspace and $A_x^-=I-A_x^+$ on the $-1$ subspace. We take the projectors $A_x^+$ to be one-dimensional real-valued projectors, parameterized by the unit vectors $\vec{A}_x \in \mathbb{R}^N$, i.e.,  $A_x^+=|a_x\rangle\langle a_x|$, with $|a_x\rangle=\sum_{i=1}^N \vec{A}_{xi}|i\rangle$. The measurement operators for Bob are defined in the same way with unit vectors $\vec{B}_y \in \mathbb{R}^N$.

The measurements of Alice are then defined by
\begin{equation}
\begin{tabular}{lclccccc}\label{SettA}
 $\vec A_1$ & $=$ & $(0,$ & $...$ & $0,$ & $0,$ & $0,$ & $1)$\\
 $\vec A_2$ & $=$ & $(0,$ & $...$ & $0,$ & $-p_2,$ & $\frac{p_1}{N-1},$ & $p_0)$\\
 $\vec A_3$ & $=$ & $(0,$ & $...$ & $-p_3,$ & $\frac{p_2}{N-2},$ & $\frac{p_1}{N-1},$ & $p_0)$\\
 $ \quad \,\,\,$ & $\vdots$ & & &  & & & \\
 $\vec A_{N-1}$ & $=$ & $(-p_{N-1},$ & $...$ & $\frac{p_3}{N-3},$ & $\frac{p_2}{N-2},$ & $\frac{p_1}{N-1},$ & $p_0)$\\
 $\vec A_N$ & $=$ & $(p_{N-1},$ & $...$&  $\frac{p_3}{N-3},$ & $\frac{p_2}{N-2},$ & $\frac{p_1}{N-1},$ & $p_0)$\\
\end{tabular}
\end{equation}
where $p_0^2=\frac{1}{N}$, $p_1^2=\frac{N-1}{N}$, and $p_{k+1}^2 =(1-\frac{1}{(N-k)^2})p_k^2$ for $k\geq 1$. The measurements of Bob are defined by
\begin{equation}
\begin{tabular}{lclccccc}\label{SettB}
 $\vec B_1$ & $=$ & $(0,$ & $...$ &  $0,$ & $0,$ & $-q_1,$ & $q_0)$\\
 $\vec B_2$ & $=$ & $(0,$ & $...$ &  $0,$ & $-q_2,$ & $\frac{q_1}{N-1},$ & $q_0)$\\
 $\vec B_3$ & $=$ & $(0,$ & $...$ &  $-q_3,$ & $\frac{q_2}{N-2},$ & $\frac{q_1}{N-1},$ & $q_0)$\\
 $ \quad \,\,\,$& $\vdots$ & & &  & & & \\
 $\vec B_{N-1}$ & $=$ & $(-q_{N-1},$ & $...$ & $\frac{q_3}{N-3},$ & $\frac{q_2}{N-2},$ & $\frac{q_1}{N-1},$ & $q_0)$\\
 $\vec B_N$ & $=$ & $(q_{N-1},$ & $...$&  $\frac{q_3}{N-3},$ & $\frac{q_2}{N-2},$ & $\frac{q_1}{N-1},$ & $q_0)$\\
\end{tabular}
\end{equation}
where $q_1^2+q_0^2=1$, and $q_{k+1}^2 =(1-\frac{1}{(N-k)^2}) q_k^2$ for $k\geq 1$.

The probabilities entering (\ref{ineq2}) can now be readily calculated. We find
\ba\label{probs}
&&P(A_1) =\epsilon^2 \nonumber\\
&&P(B_y) =\frac{1-\epsilon^2}{N-1}(1-q_0^2)+\epsilon^2 q_0^2 \quad \textrm{ for } \,2 \leq y \leq N \nonumber\\
&&P(A_1,B_y) = \epsilon^2 q_0^2 \quad \textrm{  for  } \, 1\le y\le N \\
&&P(A_x,B_x) = \left(\sqrt\frac{1-\epsilon^2}{N-1} p_1 q_1  + \epsilon p_0 q_0 \right)^2 \textrm{  for  } \, x\ge 2\nonumber \\
&&P(A_x,B_y) = \left(\sqrt\frac{1-\epsilon^2}{N-1} \frac{p_1 q_1}{1-N} + \epsilon p_0 q_0 \right)^2  \textrm{  for  } \, x>y\ge 1\nonumber
\ea

In order to maximize the quantum violation of the inequality \eqref{ineq2}, we choose the free parameter $\epsilon$ defining the quantum state (\ref{state}) as $\epsilon^2 = \frac{1-q_0^2}{1+[(N-1)^2-1]q_0^2}$, so that the joint probabilities $P(A_x,B_y)$ with $x>y \geq 1$ cancel out. Note that $0\leq\epsilon\leq 1$ for $0\leq q_0\leq 1$. We then find that
\ba I_{NN22}(\eta_{B}) = \epsilon^2 \left( -\frac{1}{\eta_{B}}+ q_0^2 N \right). \ea
This quantity is positive if $\eta_{B}>\frac{1}{Nq_0^2}$. Therefore, the inequality $I_{NN22}(\eta_{B})\leq0$ can be violated for any value of $\eta_{B}>\frac{1}{N}$ by taking $q_0$ sufficiently close to 1.

In the limit $q_0 \rightarrow 1$, we get $\epsilon \rightarrow 0 $. Thus the state providing the lowest detection efficiency is arbitrarily close to a maximally entangled state of dimension $N-1$. In particular in the qubit case ($N=2$), it becomes close to a separable state, analogously to the results in \cite{Eberhard,CabelloDetLoop,AsymDetLoop}.

Note that while the measurement settings (4,5) are well-adapted for partially entangled states ($\epsilon \rightarrow 0$), they do not provide the maximal violation for all values of $\epsilon$. It would be interesting to find a construction that is optimal for any degree of entanglement.
For the case $N=3$, we optimized numerically the unit vectors $\vec{A}_x$ and $\vec{B}_y$ to find the optimal threshold efficiency $\eta_{B}$ as a function of the degree of entanglement $\epsilon$ (see Fig.~1). Similarly to results in the qubit case \cite{Eberhard,CabelloDetLoop,AsymDetLoop}, $\eta_B$ decreases with $\epsilon$. We also investigated the influence of background noise, by replacing the pure state $\ket{\psi_\epsilon}$ in Eq.~\eqref{state} by a mixed state of the form $\rho = (1-p) \ket{\psi_\epsilon}\bra{\psi_\epsilon}+p\openone/N^2$, where $p$ is the amount of white noise. For small values of $\epsilon$, $\eta_{B}$ becomes quite sensitive to noise, due to the fact that the violation of the inequality $I_{3322}(\eta_{B})\leq 0$ becomes small.


\begin{figure}[t]
  \includegraphics[width=0.79\columnwidth]{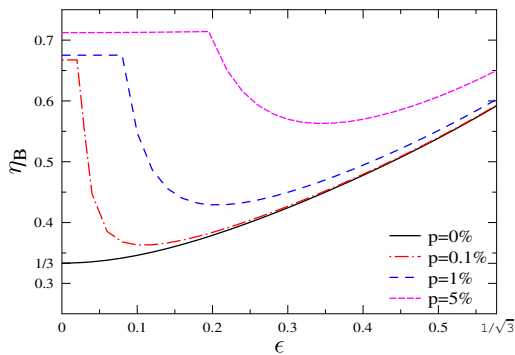}
  \caption{Asymmetric Bell test with qutrits. Threshold efficiency $\eta_{B}$ as a function of the degree of entanglement $\epsilon$ for different values of background noise $p$. For $p=0$, the efficiency tends to $33.3\%$ when $\epsilon\rightarrow 0$, and is equal to $66.7\%$ at the value $\epsilon=0$ (i.e. for maximally entangled qubits). For $p>0$, the curve exhibits a plateau when the efficiency becomes equal to the one given by maximally entangled qubits.}
\end{figure}\label{fig:Asym}

\emph{Dimension witnesses.}
For the case $N=3$, we have strong numerical evidence that the inequality $I_{3322}(\eta_B)\leq 0$ cannot be violated by performing measurement on qubits if $\eta_{B}\leq 0.428$. Therefore the inequality $I_{3322}(\eta_{B}=0.428)\leq0$ is a dimension witness \cite{dimH}: its violation guarantees that qutrits (at least) have been prepared. More generally, we conjecture that the inequalities $I_{NN22}(\eta_B)\leq 0$ with $\eta_B\simeq 1/N$ can only be violated by states of local dimension $N$, and thus are $N$-dimensional witnesses.

\emph{Symmetric case.} Here we show that the threshold detection efficiency can also be significantly lowered using low dimensional qudits. Specifically, we consider the case $N=4$ and the Bell inequality $I_{4422}^4\leq 0$ introduced in Ref. \cite{I4422}, which can be re-written as
\ba\label{ineqsym}\nonumber I_{4422}^4 &=& I_{CH}^{(1,2;1,2)}+I_{CH}^{(3,4;3,4)}-I_{CH}^{(2,1;4,3)}-I_{CH}^{(4,3;2,1)} \\ & & - P(A_2)- P(A_4)- P(B_2)- P(B_4) \ea
with $I_{CH}^{(i,j;m,n)}\equiv P(A_{i},B_{m})+P(A_{j},B_{m})+P(A_{i},B_{n})-P(A_{j},B_{n})-P(A_{i})-P(B_{m})$.  To take into account inclusive events we choose that Alice and Bob output ``-1" in case of non-detection. The probabilities are thus modified according to $P(A_x,B_y)\rightarrow \eta^2 P(A_x,B_y)$, $P(A_x)\rightarrow \eta P(A_x)$, $P(B_y)\rightarrow \eta P(B_y)$. Inserting these values in (\ref{ineqsym}), we obtain, similarly to the asymmetric case, a modified (efficiency dependent) Bell inequality $I_{4422}^4(\eta)\leq 0$.

\begin{figure}[t]\label{fig:Sym1}
  \includegraphics[width=0.82\columnwidth]{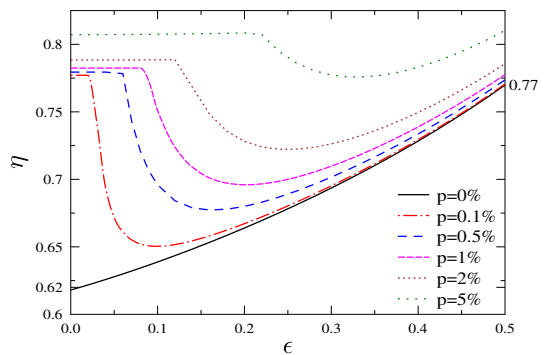}
  \caption{Symmetric Bell test with 4-dimensional systems. Threshold efficiency $\eta$ as a function of the degree of entanglement $\epsilon$ for different values of background noise $p$. For maximally entangled states ($\epsilon=\frac{1}{2}$) an efficiency of $76.98\%$ can be tolerated; for very partially entangled states ($\epsilon \rightarrow 0$), the efficiency drops down to $61.8\%$.}
\end{figure}

Next we consider entangled states of the form \eqref{state} with $N=4$ (entangled ququarts), and measurement operators parameterized as previously by unit vectors $\vec{A}_x$ and $\vec{B}_y$ (see
Appendix A for details). In the limit $\epsilon\rightarrow 0$, we
show that the inequality $I_{4422}^4(\eta)\leq 0$ is violated
if $\eta>(\sqrt{5}-1)/2\simeq 0.618$. Using the techniques
introduced in \cite{npa}, we checked that this value is optimal
for $I_{4422}^4$, i.e. that no violation is possible if $\eta\leq
0.618$. For the maximally entangled state ($\epsilon =
\frac{1}{2}$), an efficiency of $76.98\%$ can be tolerated. Fig.~2
presents the minimal detection efficiency $\eta$ for intermediate
values of $\epsilon$, obtained by performing a numerical
optimization over the unit vectors $\vec{A}_x$ and $\vec{B}_y$.
These results represent a significant improvement over the best
values known so far for small systems, namely $79.39\%$ for
maximally entangled ququarts \cite{zoology}, and $66.67\%$ for
weakly entangled qubits \cite{Eberhard}. In Fig.~3, we
analyze the influence of background noise. For experimentally
realistic values of the background noise ($p<2\%$), our model can
tolerate efficiencies about $5-8\%$ lower compared to the CHSH inequality.

\emph{Experimental perspectives.} Going beyond qubits, but nevertheless using low dimensional quantum systems, allowed us to significantly lower the detection efficiencies required to close the detection loophole in Bell experiments. Since we used Bell inequalities with few settings and simple measurements, we believe that our findings are relevant from experimental perspectives.

In the asymmetric case, related to atom-photon entanglement, we showed that an efficiency arbitrarily close to $1/N$ can be tolerated for the photon, using $N$-dimensional quantum states. Therefore, the (typically) low photon detection efficiency, usually the main experimental limitation, can be compensated by increasing the Hilbert space dimension, which appears feasible. Notably, qutrit entanglement has been recently observed between an atomic ensemble and a photon \cite{inoue09}. Since the experiment \cite{inoue09} used orbital angular momentum (OAM) as degree of freedom, it could in principle be performed with higher dimensional systems as well.

\begin{figure}[t!]
  \includegraphics[width=0.79\columnwidth]{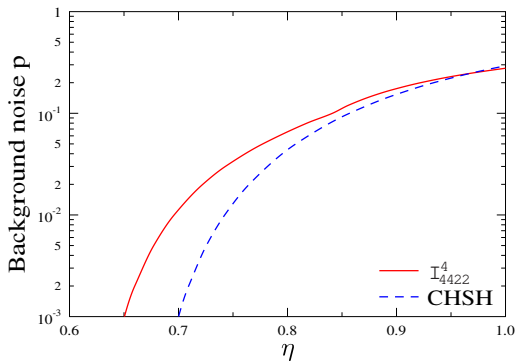}
  \caption{Minimal detection efficiency $\eta$ for a given amount of noise $p$. The curve for CHSH is Eberhard's result \cite{Eberhard}. For a realistic amount of background noise ($p<2\%$), our models can tolerate efficiencies $5-8\%$ lower than CHSH.}
\end{figure}\label{fig:Sym2}

For the symmetric case, we showed that efficiencies as low as $61.8\%$ can be tolerated using four-dimensional states, thus providing a significant improvement over CHSH -- the best inequality known so far. This appears promising from the perspective of photon experiments, in which higher dimensional entanglement has been demonstrated using OAM \cite{OAM}, time-bins \cite{thew}, multipath entanglement \cite{mpe}, and hyperentanglement \cite{kwiat97}, i.e. photons entangled in several degrees of freedom \cite{hyperentanglement}. Hyperentanglement appears here particularly relevant, since four-dimensional entangled states can be conveniently obtained using photons entangled in both polarization and mode \cite{hyperMartini}. Partially entangled states of the form \eqref{state}, as well as the required measurements, can be implemented using standard equipment (see Appendix B). Moreover, since the measurements involved in our model have binary outcomes, only one detector is required on each side. Considering a realistic background noise of $1\%$ (i.e. visibilities of $99\%$), the required overall efficiency, including transmission from the source to Alice (or Bob) and photon detection, is $\sim 69\%$ (see Fig. 3). For a measurement time of 100ns, the distance between the
source and Alice (Bob) should be of the order of 50 meters,
to ensure space-like separation. On such distance, a transmission
of the order of $80\%$ can be expected, which imposes a photon detection efficiency of at least $86\%$. This is a demanding feature, but efficiencies of the order of $90\%$ have already been reported \cite{detectors1}. Thus, overall, the perspective of implementing a loophole-free experiment using photon hyperentanglement seems very promising.

\emph{Acknowledgements.} We thank N. Gisin, E. Pomarico, J. Rarity, V. Scarani, and H. Zbinden for helpful discussions, and acknowledge financial support from the Swiss National Science Foundation, the Swiss NCCR Quantum Photonics, the European ERC-AG QORE, and from a J\'anos Bolyai Programme of the Hungarian Academy of Sciences.

\subsection{Appendix A: Symmetric case}
Here we give a detailed derivation of our construction for the symmetric case.
We use the four-setting Bell inequality $I_{4422}^4$ of Ref. \cite{I4422}, and show that it allows to improve over CHSH, both for maximally and for partially entangled states. As a starting point we present a simple construction to recover Eberhard's result for the CHSH inequality \cite{Eberhard}.

\emph{An analytic construction for Eberhard's result.}
Here we write the CHSH inequality in the Clauser-Horne (CH) form:
\ba\label{ch}
I_{CH}&\equiv& P(A_1,B_1) + P(A_1,B_2) + P(A_2,B_1)\nonumber\\ & &- P(A_2,B_2)
- P(A_1) - P(B_1)\leq 0 \ea
We choose that both Alice and Bob always output ``-1" in case of non-detection, so that all measurement outputs are still binary. The measurement probabilities are modified according to
\ba
P(A_x,B_y)&\rightarrow& \eta^2 P(A_x,B_y)\nonumber\\
P(A_x)&\rightarrow& \eta P(A_x)\nonumber \\
P(B_y)&\rightarrow& \eta P(B_y)
\ea
Introducing these expressions in (\ref{ch}) and dividing by $\eta$, we obtain a modified (efficiency-dependent) inequality $I_{CH}(\eta)\leq0$, where
\ba\label{CH}\nonumber I_{CH}(\eta)&=& P(A_1,B_1) + P(A_1,B_2) + P(A_2,B_1) \\ & &- P(A_2,B_2)
- \frac{P(A_1)}{\eta} - \frac{P(B_1)}{\eta} . \ea
We study the violation of this inequality for qubit states of the form (\ref{state}) (corresponding to $N=2$) and define the measurement settings trough the following unit vectors:
\begin{equation}
\begin{tabular}{lcclcc}
$\vec A_1=$ & ($-u,$ & $p_1$) & \,\, $\vec A_2=$ & ($v,$ & $p_2$)
\vspace{0.25cm}\\
$\vec B_1=$ & ($u,$ & $q_1$) & \,\, $\vec B_2=$ & ($-v,$ & $q_2$).\\
\end{tabular}
\label{vecCH}
\end{equation}
Note, that here $p_1$ and $p_2$ are the only variables, since $q_1=p_1$, $q_2=p_2$ and $u=\sqrt{1-p_1^2}$, $v=\sqrt{1-p_2^2}$ due to normalization.

We first consider the case of a maximally entangled pair of qubits ($\epsilon=1/\sqrt{2}$). Then, by setting $u=\sin(\pi/16)$ and $v=\sin(3\pi/16)$, the modified CH inequality~(\ref{CH}) is violated if $\eta>2/(1+\sqrt 2)\sim 0.828$. For a partially entangled pair of qubits in the limit $\epsilon\rightarrow 0$, by choosing $u=0$, $v=\epsilon^{1/2}$, the CH inequality~(\ref{CH}) is violated if $\eta>2/3$, which is optimal as shown in Ref. \cite{mp03}.

\emph{$I_{4422}^4$ inequality with four-dimensional systems.} Now
let us move to the four-setting inequality $I_{4422}^4\leq 0$
\cite{I4422}, where \ba\nonumber I_{4422}^4 &\equiv&
I_{CH}^{(1,2;1,2)}+I_{CH}^{(3,4;3,4)}-I_{CH}^{(2,1;4,3)}-I_{CH}^{(4,3;2,1)}
\\ & & - P(A_2)- P(A_4)- P(B_2)- P(B_4) \ea and
$I_{CH}^{(i,j;m,n)}\equiv
P(A_{i},B_{m})+P(A_{j},B_{m})+P(A_{i},B_{n})-P(A_{j},B_{n})-P(A_{i})-P(B_{m})$.
Again, in case of non-detection both Alice and Bob will output
``-1". Proceeding as above, we obtain a modified version of the
inequality $I_{4422}^4(\eta)\leq0$, where
\ba\nonumber\label{I4422mod} I_{4422}^4(\eta) &\equiv&
I_{CH}^{(1,2;1,2)}(\eta)+I_{CH}^{(3,4;3,4)}(\eta)\\ & &
-I_{CH}^{(2,1;4,3)}(\eta)-I_{CH}^{(4,3;2,1)}(\eta) \\\nonumber & &
- \left[ P(A_2)+ P(A_4)+ P(B_2)+ P(B_4) \right]/\eta. \ea Here we
have that $I_{CH}^{(i,j;m,n)}(\eta)\equiv
P(A_{i},B_{m})+P(A_{j},B_{m})+P(A_{i},B_{n})-P(A_{j},B_{n})-
\left[ P(A_{i})+P(B_{m})\right] / \eta $. Note that the maximum
quantum value of the Bell expression $I_{4422}^4$ is obtained with
degenerate measurements and using real qubits \cite{pal09}.

Let us define Alice's and Bob's measurement settings with the
following four-dimensional unit vectors:
\begin{equation}
\begin{tabular}{lccclccc}
$\vec A_1=$ & ($-u,$ & $-u,$ & $\vec p_1$) & \,\, $\vec A_3=$ & ($u,$ & $u,$ & $\vec p_1$)\\
$\vec A_2=$ & ($-v,$ & $v,$ & $\vec p_2$) & \,\, $\vec A_4=$ & ($v,$ & $-v,$ & $\vec p_2$)
\vspace{0.25cm}\\
$\vec B_1=$ & ($-u,$ & $u,$ & $\vec q_1$) & \,\, $\vec B_3=$ & ($u,$ & $-u,$ & $\vec q_1$)\\
$\vec B_2=$ & ($-v,$ & $-v,$ & $\vec q_2$) & \,\, $\vec B_4=$ & ($v,$ & $v,$ & $\vec q_2$)\\
\end{tabular}
\label{vecI4422}
\end{equation}
where $\vec p,\vec q$ are two-dimensional vectors, $\vec
p_i=(p_{i1},p_{i2})$, $\vec q_i=(q_{i1},q_{i2})$ for $i=1,2$. Note
that vectors $\vec A,\vec B$ above can be built up from the ones
we used previously for the CH inequality \eqref{vecCH}. In
particular, the last two columns of $\vec A_x$, $x=1,2$ ($\vec
A_x$, $x=3,4$) and of $\vec B_y$, $y=3,4$ ($\vec B_y$, $y=1,2$)
correspond Alice's vectors (Bob's vectors) in (\ref{vecCH}) except
that now $p,q$ refer to vectors. The first columns of the vectors,
on the other hand, are filled up by alternating the components $u$
and $v$ either with only plus signs or only with minus signs.
Using numerical optimization, we find the optimum parameters for a
pair of maximally entangled four-dimensional states (ququarts)
($\lambda_i=1/2$, $i=1,2,3,4$) to be given by
\begin{equation}
\begin{tabular}{lcc}
$\vec p_1=$ & ($0.9159,$ & $0.0499$)\\
$\vec p_2=$ & ($0.5625,$ & $-0.3035$)
\vspace{0.25cm}\\
$\vec q_1=$ & ($0.9159,$ & $-0.0499$)\\
$\vec q_2=$ & ($0.5625,$ & $0.3035$).\\
\end{tabular}
\label{vecI4422pq}
\end{equation}
Since $\vec A_x$ and $\vec B_y$ are unit vectors, $u$ and $v$ are
completely determined by the vectors $\vec p$ and $\vec q$ defined
above. They yield the threshold value $\eta=0.7698$ for the
inequality \eqref{I4422mod}.

Now let us move to partially entangled ququarts. Considering the
limit of small $\epsilon$, we set $u=\sqrt{3/8}(\sqrt 5
-1)\epsilon$, $v=1/2$ and $\vec p_1=\vec q_1=(0,1)$, $\vec
p_2=(\delta, 1/\sqrt 2)$, $\vec q_2=(-\delta, 1/\sqrt 2)$ with
$\delta=(3/4)^{1/4}\epsilon^{1/2}$. This yields the threshold
$\eta_s>(\sqrt 5 -1)/2\sim 0.618$ in the limit
$\epsilon\rightarrow 0$. Furthermore, one can check numerically
(using a program similar to Sec.~II.D of \cite{zoology}) that the
above threshold efficiencies are the lowest possible ones allowed
by the construction \eqref{vecI4422} both for the maximally
entangled and for partially entangled ququarts.

\subsection{Appendix B: Optical implementation using hyperentanglement}

We now show how our construction for the symmetric case can be implemented using photon hyperentanglement. We first show how partially entangled states of the form \eqref{state} can be created; then we show how to perform the required measurements. Importantly, both stages can be implemented using only linear optics (except for non-linear crystals), and seem feasible using current technology.
\begin{figure}[b]
\label{fig:StatePrep}
  \includegraphics[width=0.8\columnwidth]{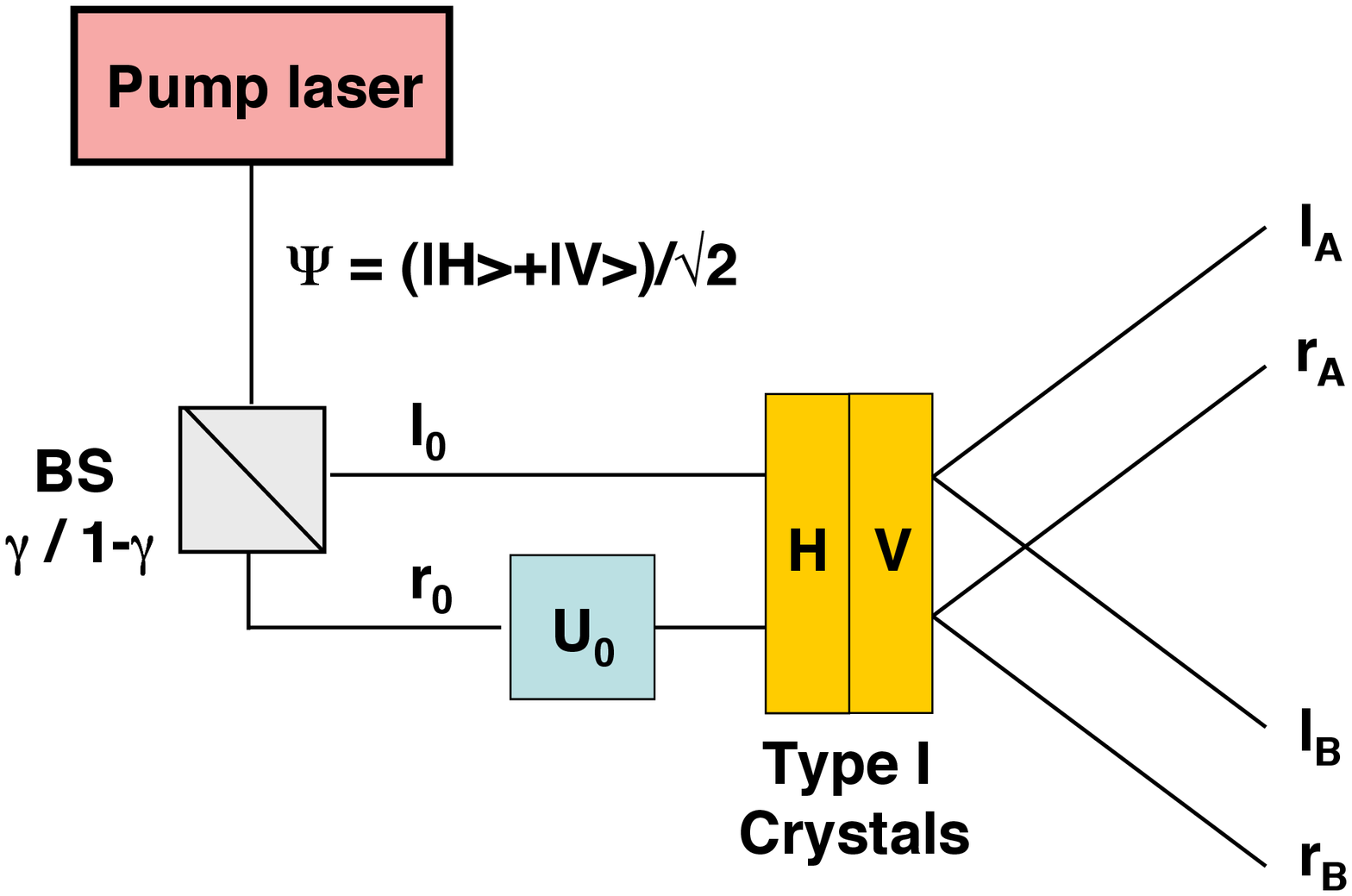}
  \caption{Scheme for preparing partially entangled quqarts state using hyperentanglement.}
\end{figure}

\emph{State preparation.} We focus on hyperentanglement using polarization and mode entanglement \cite{hyperMartini}. The scheme for the state preparation is sketched in Fig.~4, and was suggested to us by John Rarity. The beam of a pump laser is polarized in the state $\ket{\psi}=\frac{1}{\sqrt{2}}(\ket{H}+\ket{V})$ and then splitted in two modes ($\ket{l_0}$ and $\ket{r_0}$) by an unbalanced beam splitter (BS), with reflexion probability $\gamma$ and transmission probability $(1-\gamma)$. Equivalently, the unbalanced BS can be replaced by a 50/50 BS followed by an attenuator on one of the modes. The polarization state of mode $\ket{r_0}$ is then rotated to an arbitrary linear polarization $\ket{\phi}=\cos{\theta}\ket{H}+\sin{\theta}\ket{V}$ by an appropriate combination of waveplates, represented on Fig.~5 by the unitary operation $U_0$. Thus before parametric down-conversion, the state is given by
\ba \sqrt{\gamma} \ket{\psi} \ket{l_0} +   \sqrt{1-\gamma}  \ket{\phi} \ket{r_0} \ea
Then, parametric down-conversion is performed using two orthogonal type I non-linear crystals, thus resulting in the state
\ba & &\sqrt{\frac{\gamma}{2}} \left( \ket{H,H}+\ket{V,V} \right) \otimes \ket{l_A,l_B} \\\nonumber &+&  \sqrt{1-\gamma} \left( \cos{\theta}\ket{H,H}+\sin{\theta}\ket{V,V} \right) \otimes \ket{r_A,r_B}\ea

By adjusting the parameters such that $\epsilon = \sqrt{1-3\gamma/2}$ and $\cos{\theta}=\sqrt{\frac{\gamma}{2(1-\gamma)}}$, we obtain the desired partially entangled state of two ququarts, i.e.

\ba\label{state4} \ket{\psi_{\epsilon}} = \sqrt{\frac{1-\epsilon^2}{3}} \left[ \ket{11}+\ket{22}+\ket{33} \right] + \epsilon \ket{44} \ea
where we have used the following encoding: $\ket{1}\equiv \ket{H}\ket{l_A}$, $\ket{2}\equiv \ket{V}\ket{l_A}$, $\ket{3}\equiv \ket{H}\ket{r_A}$, and $\ket{4}\equiv \ket{V}\ket{r_A}$ for Alice, and similar definitions for Bob.

Importantly, note that this scheme does not require post-selection.

\emph{Measurements.} All measurements required in our model are projectors on four-dimensional states of the form:
\ba \ket{\Phi} = \sum_{j=1}^4 a_j\ket{j} \ea where $\sum_{j=1}^4 |a_j|^2 =1$. Here all coefficients can be taken to be real, therefore we can rewrite $a_1=\alpha \cos{\theta_1}$, $a_2=\alpha \sin{\theta_1}$, $a_3=\beta \cos{\theta_2}$, and $a_4=\beta \sin{\theta_2}$, where $\alpha^2+\beta^2=1$. Thus the projectors can now be conveniently rewritten as
\ba \ket{\Phi} = \alpha \ket{\phi_1}\ket{l} + \beta \ket{\phi_2}\ket{r} \ea where $\ket{\phi_j}= \cos{\theta_j}\ket{H}+\sin{\theta_j}\ket{V}$ are normalized polarization states for each mode $l$ and $r$.

\begin{figure}[h!]
\label{fig:Meas}
  \includegraphics[width=0.8\columnwidth]{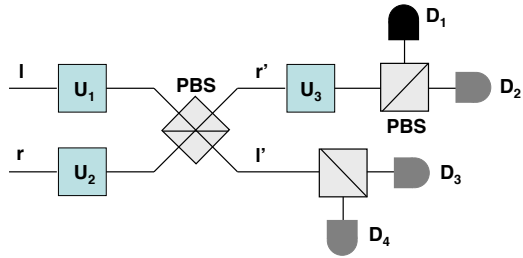}
  \caption{Setup for implementing the four-dimensional projection. Since we consider two-outcome measurements, only one detector is needed.}
\end{figure}
This measurement can be implemented using the simple setup described in Fig.~5. First, a rotation of the polarization must be performed on each mode. The corresponding unitaries $U_1$, $U_2$ must be such that $U_1 \phi_1 = \ket{H}$ and $U_2 \phi_2 = \ket{V}$. Then the PBS implements the operation:

\ba\nonumber \ket{H}\ket{l} \rightarrow \ket{H}\ket{l'} \quad &,& \quad \ket{H}\ket{r} \rightarrow \ket{H}\ket{r'} \\ \ket{V}\ket{l} \rightarrow \ket{V}\ket{r'} \quad &,& \quad\ket{V}\ket{r} \rightarrow -\ket{V}\ket{l'} \ea The PBS maps the state $U_1 U_2\ket{\Phi}$ to a state of the form $(\alpha \ket{H}+\beta \ket{V})\otimes \ket{l'}$. The final polarization rotation is characterized by the unitary $U_3$, with $U_3 (\alpha \ket{H}+\beta \ket{V}) = \ket{V}$. Thus an incoming photon in the state $\ket{\Phi}$ always produces a click in detector $D_1$; any state orthogonal to $\ket{\Phi}$ does never produce a click in $D_1$.

The measurements required in our scheme can therefore be implemented using standard linear optics devices. Moreover, a single photon detector is enough, since we consider here two-outcome measurements.

\bibliographystyle{prsty}
\bibliography{C:/BIB/thesis}

\end{document}